\documentclass[%
 aip,
 sd,%
 amsmath,amssymb,
 reprint,%
]{revtex4-1}

\usepackage{graphicx}
\usepackage{subcaption}
\usepackage{multirow}
\usepackage{dcolumn}
\usepackage{bm}
\usepackage{booktabs}
\usepackage[ruled,vlined]{algorithm2e}

\begin{document}

\preprint{AIP/123-QED}

\title[Manuscript Submitted to Review of Scientific Instruments]{A Modular Magneto-Inductive Sensor for Low Vector Magnetic Field Measurements}

\author{Huan Liu}
\email{huan.liu@cug.edu.cn (Authors to whom correspondence should be addressed)}
\affiliation{School of Automation, China University of Geosciences, Wuhan, Hubei 430074, China}
\affiliation{Hubei Key Laboratory of Advanced Control and Intelligent Automation for Complex Systems, Wuhan, Hubei 430074, China}
\affiliation{Engineering Research Center of Intelligent Technology for Geo-Exploration, Ministry of Education, Wuhan, Hubei 430074, China}
\affiliation{School of Engineering, University of British Columbia Okanagan Campus, Kelowna, BC V1V 1V7, Canada}

\author{Xiaobin Wang}
\affiliation{School of Automation, China University of Geosciences, Wuhan, Hubei 430074, China}
\affiliation{Hubei Key Laboratory of Advanced Control and Intelligent Automation for Complex Systems, Wuhan, Hubei 430074, China}
\affiliation{Engineering Research Center of Intelligent Technology for Geo-Exploration, Ministry of Education, Wuhan, Hubei 430074, China}

\author{Changfeng Zhao}
\affiliation{School of Automation, China University of Geosciences, Wuhan, Hubei 430074, China}
\affiliation{Hubei Key Laboratory of Advanced Control and Intelligent Automation for Complex Systems, Wuhan, Hubei 430074, China}
\affiliation{Engineering Research Center of Intelligent Technology for Geo-Exploration, Ministry of Education, Wuhan, Hubei 430074, China}

\author{Zehua Wang}
\affiliation{School of Automation, China University of Geosciences, Wuhan, Hubei 430074, China}
\affiliation{Hubei Key Laboratory of Advanced Control and Intelligent Automation for Complex Systems, Wuhan, Hubei 430074, China}
\affiliation{Engineering Research Center of Intelligent Technology for Geo-Exploration, Ministry of Education, Wuhan, Hubei 430074, China}

\author{Jian Ge}
\affiliation{School of Automation, China University of Geosciences, Wuhan, Hubei 430074, China}
\affiliation{Hubei Key Laboratory of Advanced Control and Intelligent Automation for Complex Systems, Wuhan, Hubei 430074, China}
\affiliation{Engineering Research Center of Intelligent Technology for Geo-Exploration, Ministry of Education, Wuhan, Hubei 430074, China}

\author{Haobin Dong}
\affiliation{School of Automation, China University of Geosciences, Wuhan, Hubei 430074, China}
\affiliation{Hubei Key Laboratory of Advanced Control and Intelligent Automation for Complex Systems, Wuhan, Hubei 430074, China}
\affiliation{Engineering Research Center of Intelligent Technology for Geo-Exploration, Ministry of Education, Wuhan, Hubei 430074, China}

\author{Zheng Liu}
\affiliation{School of Engineering, University of British Columbia Okanagan Campus, Kelowna, BC V1V 1V7, Canada}

\date{\today}

\begin{abstract}
The low magnetic field measurement has been utilized since ancient times in order to find economic resources, to detect magnetic anomalies, etc. In this case, the vector magnetic survey can simultaneously obtain the modulus and direction information of the magnetic field, which can contribute to obtaining more precise information and characteristics of magnetic field resources. This paper is concerned with the potential to exploit the signals of vector magnetic field measurement with a magneto-inductive (MI) sensor. To evaluate the capability of the MI sensor, a test platform is set up and its performance including the noise floor, the resolution, the sensitivity, etc., are comprehensively characterized. Further, a comparative geomagnetic observation and magnetic anomaly detection among the proposed MI sensor, a high-precision Overhauser sensor, and a commonly used and accepted commercial MI sensor are conducted. The experimental results identify the capability of the proposed MI sensor in weak magnetic detection. 
\end{abstract}

                         
\keywords{Magnetic field, magneto-inductive, miniature sensor, vector measurement}

\maketitle

\section{Introduction}
The Earth’s magnetic field is a ubiquitous feature of the solar system, and it is also a critical factor for geophysical, magnetosphere, and heliospheric investigations~\cite{Luo-2020-J,Canciani-2017-J,Liu-2018-J1}. In recent years, since the magnetic information obtained by the scalar measurements is limited and their ability to identify weak magnetic objects is poor, measuring the vector magnetic field has begun to receive growing concern and is developing rapidly~\cite{Shen-2016-J,Liu-2020-J1}. When compared it with the scalar magnetic field measurements~\cite{Dong-2016-Journal,Denisov-2014-J}, the vector magnetic measurement can simultaneously obtain the modulus and direction information of the magnetic field, providing more information and reveal detailed features of the Earth's magnetic field not expressed in the scalar magnetic survey, which can effectively reduce the multiplicity on data inversion~\cite{Liu-2021-J,Behroozmand-2013-J,Liu-2019-J}.  

The vector magnetometers commonly used in geophysical engineering mainly include fluxgate magnetometers and superconducting magnetometers~\cite{Stele-2020-J}, in which the former is one of the most used vector instruments, with a resolution of about 0.1 nT. Generally, a fluxgate sensor is always employed to measure the ambient vector magnetic field with high precision~\cite{Ripka-2010-J,Ponder-2016-J}. However, in some scenarios, e.g., measuring the ambient surface magnetic field of a superconducting radio-frequency cavity, this approach is not available to achieve a higher spatial resolution since the fluxgate sensor's size is relatively large. For instance, a commonly used and accepted fluxgate sensor, dubbed Mag-03, which is developed by with dimensions of 32 mm x 32 mm x 225 mm~\cite{Pang-2020-J}. 

In recent years, a kind of magnetic sensor using magneto-inductive (MI) technique has gradually entered the field of weak magnetic detection~\cite{Moldwin-2017-J,Liu-2020-J}. The measurement principle of the MI sensor is conspicuously different from that of the fluxgate sensor. Because of its miniature size, the MI sensor has been commonly employed in many applications, such as position orienting, inertial navigation, security system, etc~\cite{Liao-2017-J,Choi-2013-Journal,Winslow-2012-J,Tian-2012-J,Zeng-2018-J}. In this paper, we develop a miniature MI sensor to measure the low vector magnetic field. The contributions include the following points:
\begin{enumerate}
    \item A miniature MI sensor with dimensions of 1 cm x 1 cm x 0.2 cm is developed, which is composed of three MI sensitive elements. 
    \item The electrical properties including the noise floor, resolution, sensitivity, etc., are evaluated in the laboratory. The testing results demonstrate that the proposed MI sensor can detect a low magnetic field strength as low as 8.28 nT when the sampling rate is 10 Hz, the typical noise floor is about 0.235 nT/Hz$^{1/2}$ at 1 Hz, the sensitivity is about 13 nT, and the non-linearity is no more than 3\%. 
    \item A test platform using the MI sensor is developed to conduct an outdoor contrast experiment with a high-precision Overhauser sensor and a commercial MI sensor, in which its performance on static magnetic field observation and magnetic anomaly detection are verified. 
\end{enumerate}

The rest of the paper is organized as follows. Section~\ref{sect:methodology} briefly describes the fundamental theories and characteristics of MI technique. Section~\ref{sect:experiment} presents the laboratory and field experimental results of the proposed MI sensor. This paper is concluded in Section~\ref{sect:conclusion} with the highlights of the contributions.

\section{Magneto-Inductive Sensor}
\label{sect:methodology}
\begin{figure}[htb]
\centering
\includegraphics[width=0.9\linewidth]{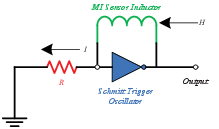}
\caption{Electrical components involved in the MI technique.}
\label{MI}
\end{figure}

The MI sensor mainly consists of a resistor-inductor circuit, in which an electronic component is sensitive to external radiation in traditional fluxgate magnetometer designs~\cite{Leuzinger-2010-J}. To be specific, as shown in Fig.~\ref{MI}. The operating of the MI sensor involves the measurement of the time it takes during the charging and discharging procedures of the inductor between an upper and lower threshold through means of a Schmitt trigger oscillator. The charge-discharge interval time is proportional to the external magnetic field strength $H$. The total magnetic field $H_t$ is composed of the external magnetic field $H$ and the generated field $kI$ of the circuit as:
\begin{equation}
H_t = H + kI
\end{equation}
where $k$ stands for the conversion factor of the sensor coil, and $I$ stands for the generated current. In this case, the Schmitt trigger would regulate the $I$ through the resistor $R$ to oscillate since the generated voltage of $R$ is over than the setting trigger value (refer to a threshold). 

\begin{figure}[htb]
\centering
\includegraphics[width=0.9\linewidth]{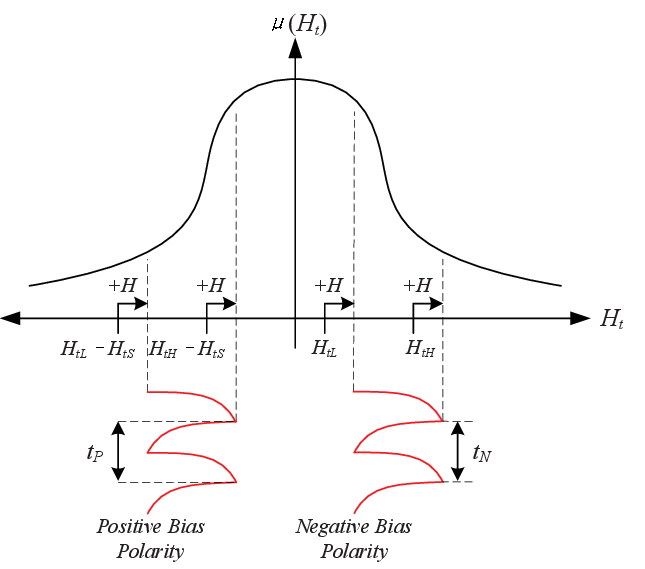}
\caption{Diagram of the induction as a function of the applied magnetic field and the traces of the oscillating current in the coils and the period for positive and negative bias polarity.}
\label{MI1}
\end{figure}

As the applied current oscillates the inductance of the circuit, the time constant changes, as shown in Fig.~\ref{MI1}, in which $\mu (H_t)$ stands for the induction curve of the sensing coils, $t_P$ and $t_N$ stand for the positive and negative bias charge and discharge time, respectively. $H_{tL}$ and $H_{tH}$ stand for the low and high charge thresholds, respectively. $H_{tS}$ represents the positive to negative bias. Generally, when there is no external magnetic field applied to the sensor, the charge and the discharge times measured as both polarities are the same, i.e., $t_P = t_N$. However, when an external field $H$ is applied, the working region along the curve will be shifted in one direction, and thus the charge and discharge times will no longer be equal, i.e., $t_P \ne t_N$. This time difference is proportional to the applied external field $H$. Hence, by measuring the time to complete a fixed number of oscillations that occur in the forward and reverse polarity directions and taking the difference between these two values, it is possible to derive the strength of the external magnetic field.

\begin{figure}[htb]
\centering
\includegraphics[width=0.86\linewidth]{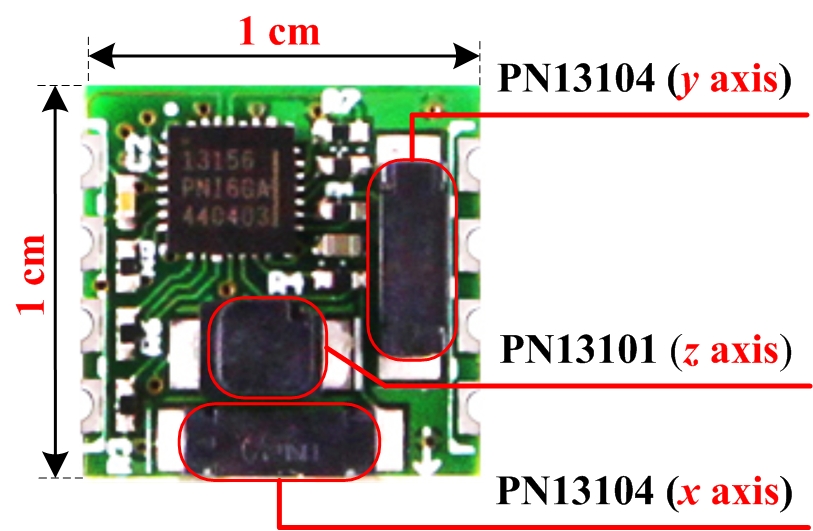}
\caption{Three-axis MI hybrid sensor.}
\label{RM3100}
\end{figure}

A MI sensor module is constructed as shown in Fig.~\ref{RM3100}, which consists of one sensor coil (PN 13101) for $z$ axis, two sensor coils (PN 13104) for $x$ and $y$ axis, respectively. Further, an application-specific integrated circuit controller is embedded in this module, which can transfer the analog data to a digital format. In this case, we can obtain the external magnetic field value through deriving the digital data to a micro-controller directly, which can avoid the additional signal conditioning module and acquisition module as that in a fluxgate sensor.

\section{Experimental results and discussion}
\label{sect:experiment}
In this section, the specifications of the proposed MI sensor including the resolution, the sensitivity, the noise floor, the stability, the linearity, and the magnetic tracking performance are evaluated by using a magnetic shielding cylinder with three layers, which can further suppress the interference of any background magnetic noise. To be specific, the remanence is less than 2 nT, the length of the homogeneity area is large than 200 mm, and below 1 kHz and over 1 kHz bands, the attenuation factor can reach 75 dB and 100 dB, respectively. The measurement range of the MI sensor is set as -800 $\mu$T to 800 $\mu$T, the digital resolution is 13.33 nT per least significant bit, and the sampling rate is 10 Hz. 

\subsection{Platform setup}
\begin{figure}[htb]
\centering
\includegraphics[width=1.0\linewidth]{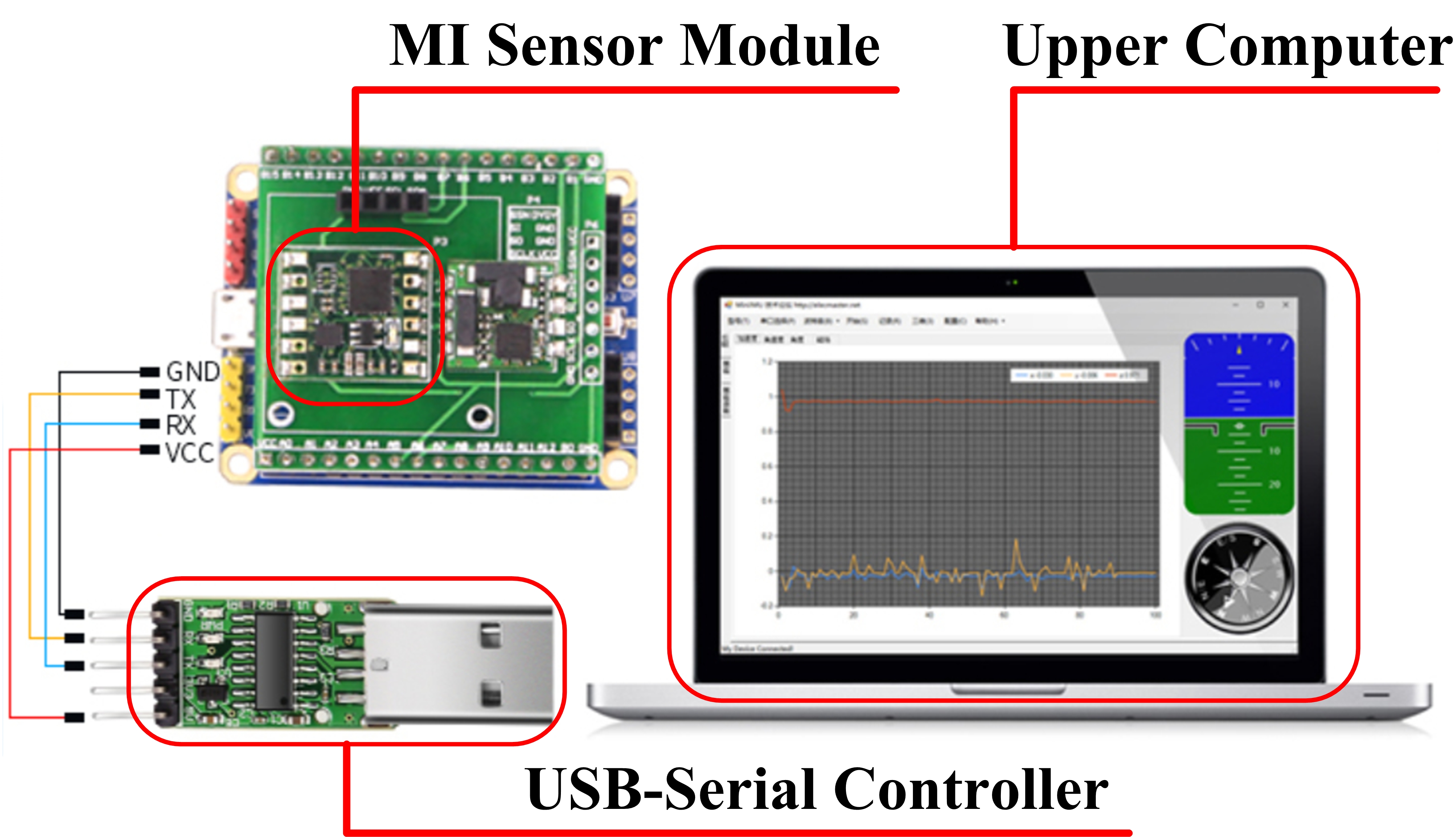}
\caption{Testing platform of the proposed MI sensor system.}
\label{Prototype}
\end{figure}

We set up a magnetic field measurement system using the proposed MI sensor as illustrated in Fig.~\ref{Prototype}. This system mainly consists of four modules, i.e., a MI sensor, a micro-controller, a universal serial bus (USB) serial controller, and an upper computer. Hence, the magnetic field data collected by the MI sensor can be transferred to the computer through the serial port, and ultimately achieve real-time monitoring. Further, a least squares based data smoothing method is employed to suppress the fluctuations of the anomalous data, further improving the signal to noise ratio (SNR) of the measured data. There are analog voltage drain drain (AVDD) and digital voltage drain drain (DVDD) in this sensor, and the AVDD and DVDD should be tied to the analog and digital supply voltages, respectively. These two powers are both 3.3 V which are yielded by two low dropout regulators and the power of the overall MI system is supplied by a 5 V battery.  

\subsection{Resolution}
\begin{figure}[htb]
\centering
\includegraphics[width=0.9\linewidth]{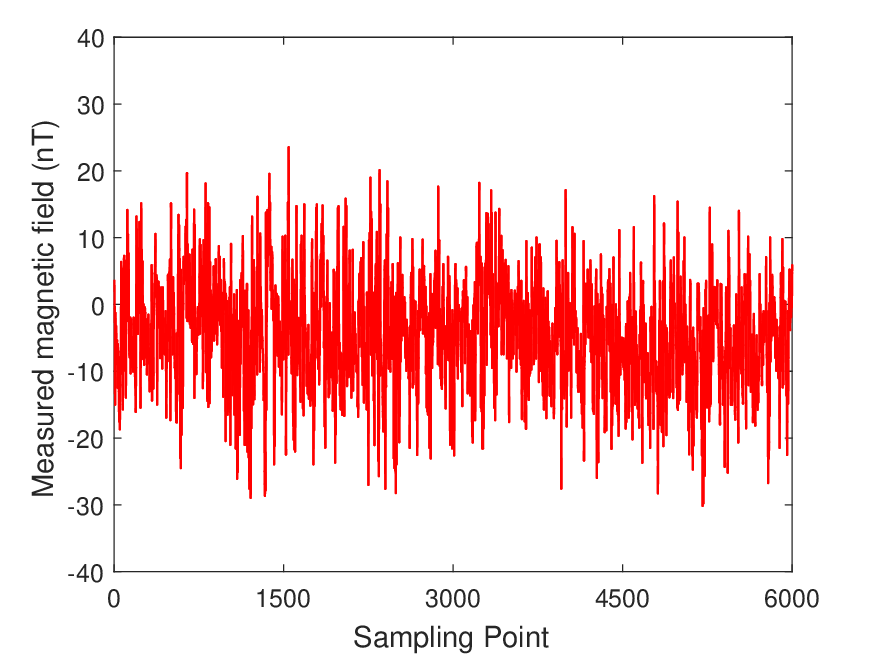}
\caption{Zero applied field measurement results.}
\label{SD}
\end{figure}

The resolution stands for the minimum detection value of the MI sensor, and the standard deviation can be adopted to evaluate the resolution~\cite{Dong-2016-Journal}. The standard deviation can be written as:
\begin{equation}
\delta = \sqrt{\frac{1}{n-1} \sum\limits_{i = 1}^n {(y_i-\overline{y_i})^2}}
\label{2}
\end{equation}
where $\delta$ stands for the standard deviation, $\overline{y_i}$ stands for the average value, and $n$ stands for the number of the collected data. Fig.~\ref{SD} shows the collected continuous data within 10 minutes. According to the equation of the standard deviation, we can calculate the resolution as about 8.28 nT. This value is lower than the digital resolution we set as 13.33 nT (each sensor reading consists of three 24 digits of data which are respectively stored in each axis registers and each number is directly proportional to the strength of the local magnetic field in the direction of the specified axis), because we implement an averaging operation for every ten data points to further improve the resolution. 

\subsection{Sensitivity}
The sensitivity of the MI sensor can be evaluated using the following equation as:
\begin{equation}
S = \frac{B_{max}-B_{min}}{\hat{B_{max}}-\hat{B_{min}}}
\end{equation}
where $B_{max}$ and $B_{min}$ stand for the maximum and the minimum magnetic field values measured by the sensor, respectively. $\hat{B_{max}}$ and $\hat{B_{min}}$ stand for the maximum and the minimum synthetic magnetic fields applied to the MI sensor which are generated by the shielding cylinder. In this case, we can calculate the sensitivity as about 13 nT.

\subsection{Noise floor}
\begin{figure}[htb]
\centering
\includegraphics[width=0.95\linewidth]{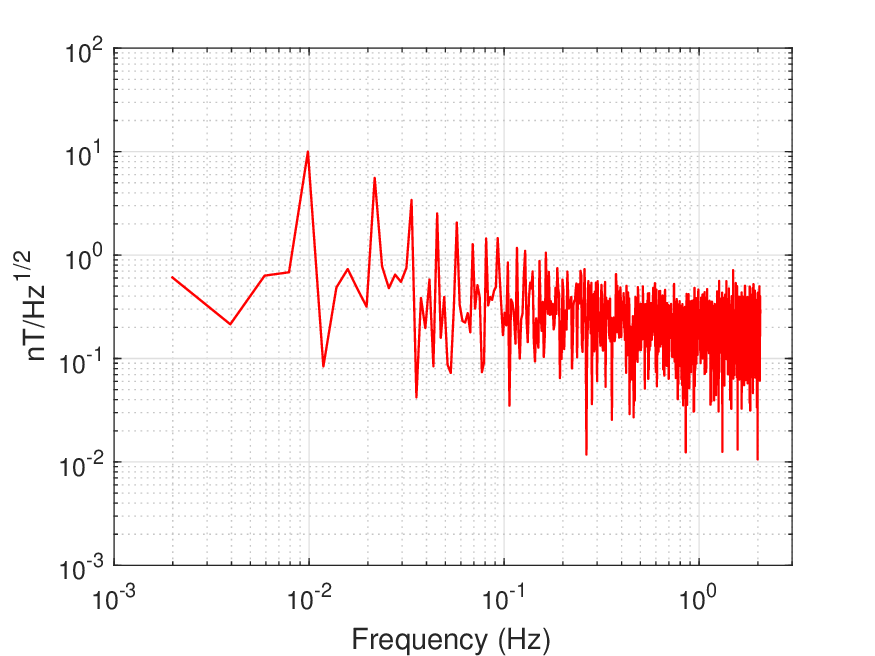}
\caption{Power spectral density of the proposed MI sensor.}
\label{Noise}
\end{figure}

Since the frequency characteristic of the noise features a 1/$f$ dependence, the power spectral density at 1 Hz is always employed to evaluate the noise floor of a sensor~\cite{Tan-2019-J}. Hence, we collect the magnetic field data consecutively within 1 hour. Fig.~\ref{Noise} shows the evaluating results of the power spectral density. We can see that the noise floor of the 1/$f$ region is relatively high, and it decreases slowly overall with up and down fluctuations as the frequency increases. As the frequency increases around 1 Hz, the total noise tends to be stable and the noise floor of the developed MI sensor is about 0.235 nT/Hz$^{1/2}$ at 1 Hz.

\subsection{Stability}
\begin{figure}[htb]
\centering
\includegraphics[width=0.95\linewidth]{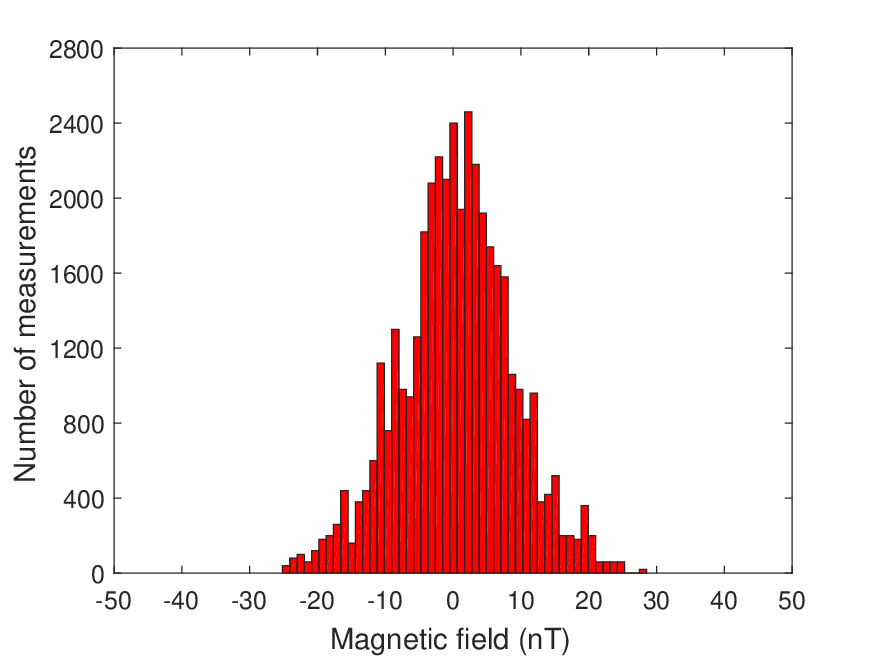}
\caption{Distribution of collected data within one hour.}
\label{Stability}
\end{figure}

The stability of the sensor represents how constant the collected data is when there is no external magnetic field, i.e., zero magnetic field. To evaluate the stability of the proposed MI sensor, the sensor is also placed in the magnetic shielding cylinder without an external magnetic field but might be accompanied by a little bit residual of the geomagnetic field. The system is operating for about 1 hour with a sampling rate of 10 Hz. Fig.~\ref{Stability} shows a histogram with the distribution of the collected data. It can be seen that the distribution is normal with an approximate symmetrical shape, and the value of the symmetrical center is about 0 nT. In addition, the randomness of the observed variations corresponds to the intrinsic noise of the sensor. 

\subsection{Linearity}
\begin{figure}[htb]
\centering
\includegraphics[width=1.0\linewidth]{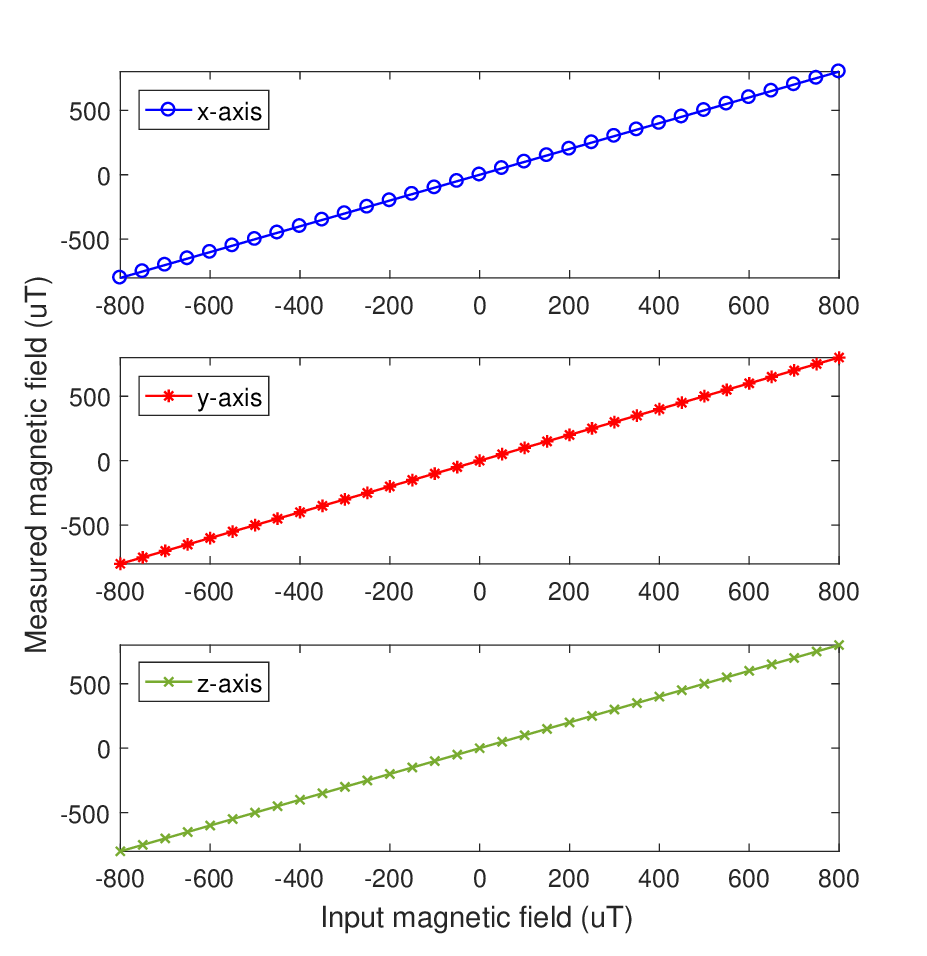}
\caption{Linearity test illustrated as the measured field vs. applied field for the MI sensor.}
\label{Geomagnetic2}
\end{figure}

The linearity of the sensor represents that when a change in the quantity being measured, how the sensor produces a proportional change throughout the whole range. To test the linearity of the MI sensor, we adopt the magnetic field shielding cylinder to generate a magnetic field varied from -800 $\mu$T to 800 $\mu$T with a step value of 50 $\mu$T, and the synthetic fields were applied aligned with the three-axis of the MI sensor, respectively. Fig.~\ref{Geomagnetic2} shows the testing results. The three curves are all close to 1, varying by less than 3\%, which implies that the proposed MI sensor remains almost linear over the whole testing range. The little deviation is generated due to the misalignment between the axis of the coil and the applied field. Since it is hard to measure this misalignment with the experimental platform, and thus the value of 3\% is regarded as an upper limit for the linearity.

\subsection{Magnetic tracking performance}
\begin{figure}[htb]
\centering
\includegraphics[width=0.95\linewidth]{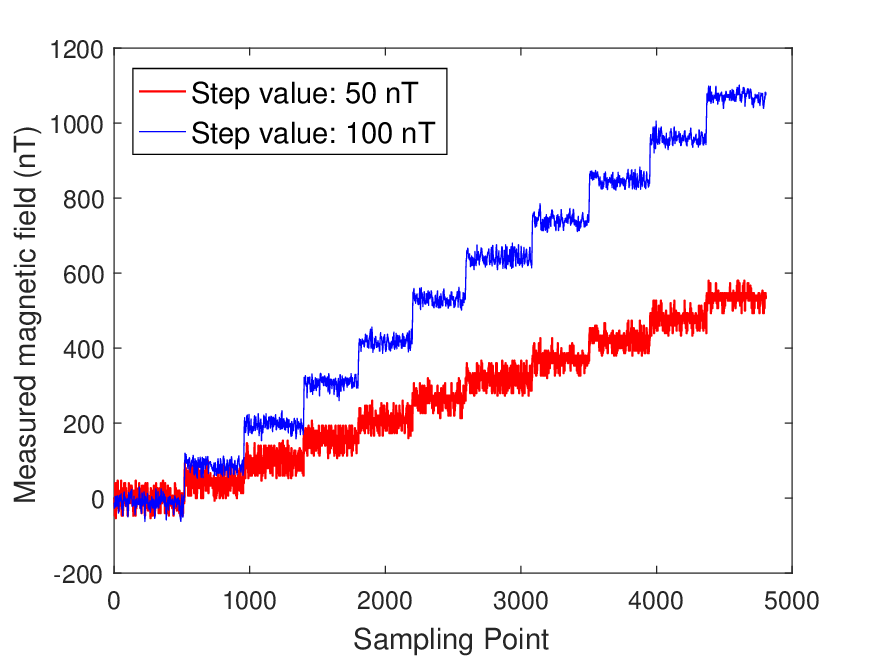}
\caption{Testing results of the magnetic tracking performance.}
\label{Resolution}
\end{figure}

The system design uses the magnetic field shielding cylinder to generate a changeable magnetic field while the step values including 50 nT and 100 nT, to identify its magnetic tracking performance. The measurement results are shown in Fig.~\ref{Resolution}. We observe that the proposed MI-based magnetometer can track the variation of 100 nT effectively, while for the 50 nT the magnetic tracking performance is less satisfactory (with major fluctuations) than that of 100 nT. In addition, there is one thing that needs to be emphasized. During the experiment, we observe that when the step value is lower than 50 nT, the magnetic tracking performance is not obvious due to the own noise properties of the MI sensor. Hence, we only depict the testing results of 100 nT and 50 nT in this paper. 

\subsection{Geomagnetic comparison test}
\begin{figure}[htb]
\centering
\includegraphics[width=0.95\linewidth]{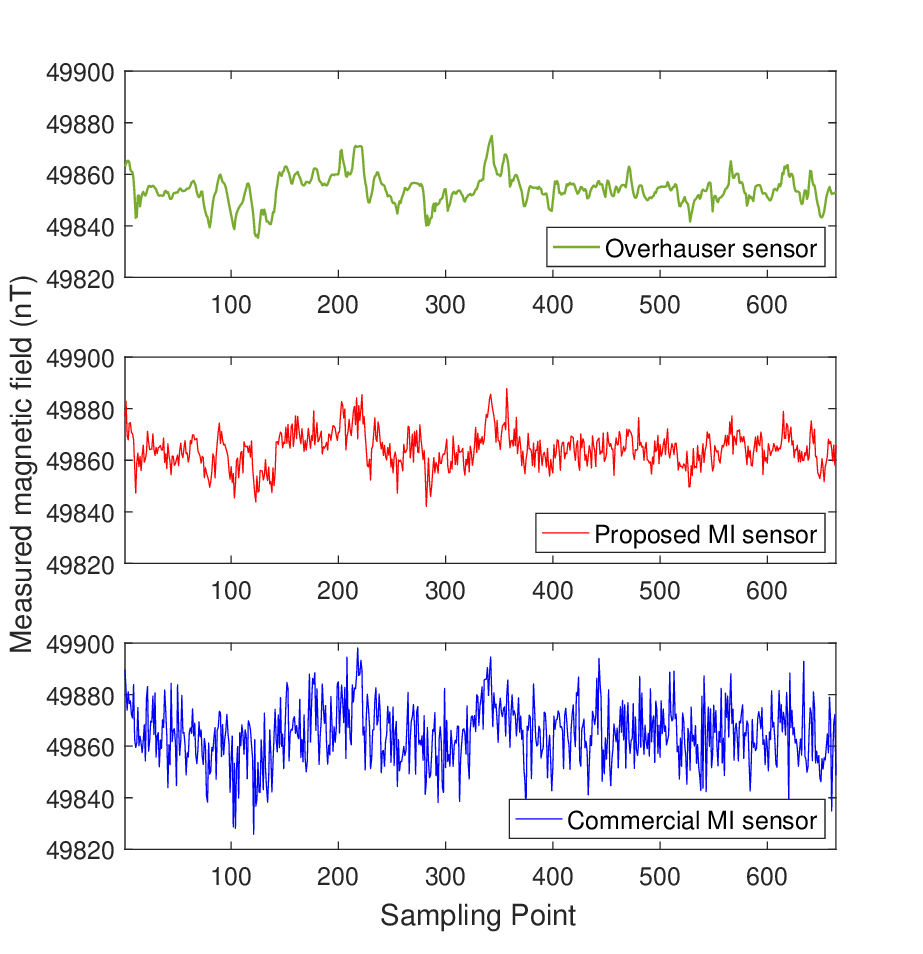}
\caption{Contrast geomagnetic observation among the commercial Overhauser magnetometer, the proposed MI sensor, and the commercial MI sensor.}
\label{Geomagnetic4}
\end{figure}

To further identify the performance of the developed MI sensor system, two comparison tests among this sensor, a high-precision Overhauser sensor, and a commonly used and accepted commercial MI sensor system were implemented to observe the magnetic field during the same period. Fig.~\ref{Geomagnetic4} illustrates the continuous geomagnetic observation results. We can see that the trend of the proposed MI sensor is approximately consistent with that of the Overhauser sensor while the data of the commercial MI sensor is with relative strong fluctuations, especially the sampling points from 50 to 100, from 300 to 400, etc. To compare the performance in a quantitative way, we first separately removed the constant baseline difference by averaging the data of each curve. Then, we excluded the large variations and evaluated the standard deviation $\delta$ for the rest data. The $\delta$s of the sampling data illustrated in Fig.~\ref{Geomagnetic4} using the Overhauser sensor, proposed MI sensor, and commercial MI sensor are 5.8290 nT, 7.9615 nT, and 11.5015 nT, respectively. The results further validate the superior performance of the proposed MI sensor.   

\begin{figure}[htb]
\centering
\includegraphics[width=0.95\linewidth]{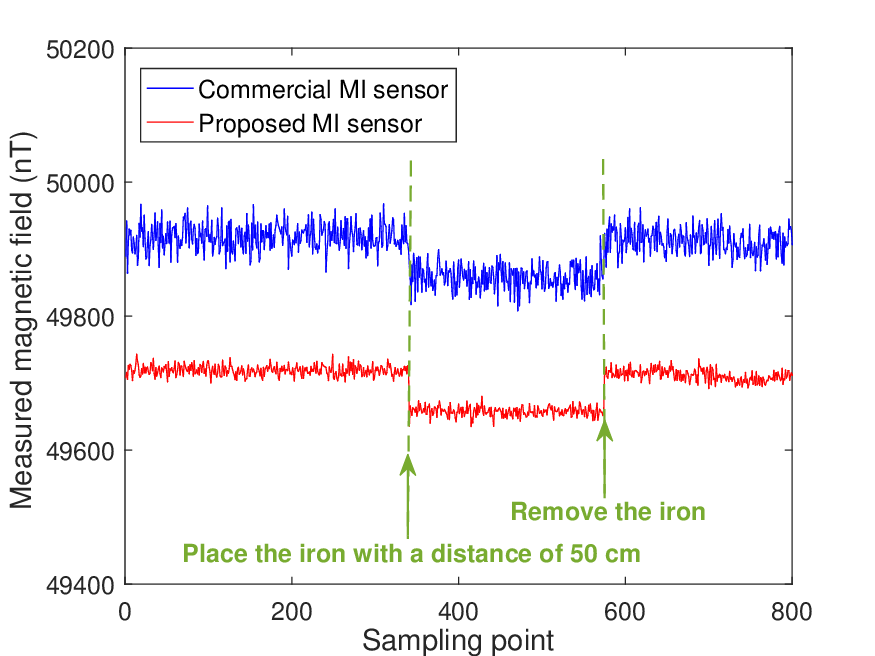}
\caption{Contrast magnetic anomaly detection between the proposed MI sensor and the commercial MI sensor.}
\label{Geomagnetic}
\end{figure}

Further, we place an iron (the shape is a cylinder and the dimension is about 80 mm x 80 mm x 200 mm) near the aforementioned two MI sensors with a distance of 50 cm, to verify the capability for magnetic anomaly detection. The local magnetic field strength is less than 50000 nT and the magnetic field measurement results are shown in Fig.~\ref{Geomagnetic}. 

Firstly, the trends of the geomagnetic field intensity measured by these two instruments are basically the same although not exactly. The main reason is that in order to avoid magnetic interference, the two magnetometers are placed 5 m far away from each other, and the different sensor positions will generate a magnetic field gradient. Besides, the fixed magnetic anomalies of buildings, vehicles, and cables in the test environment also generate magnetic field gradients, but they do not affect the contrast between the two sensors. Further, under the influence of moving pedestrians and vehicles in the test environment, the external magnetic interference cannot be completely avoided. However, the fluctuation of the magnetic field value measured by the proposed MI sensor is significantly lower than that of the commercial MI sensor (the $\delta$s of the sampling data from 350th to 550th  illustrated in Fig.~\ref{Geomagnetic} using the proposed MI sensor and commercial MI sensor are 7.8512 nT and 11.8104 nT, respectively). The consistent variation trend of the commercial MI sensor and the proposed MI sensor indicates that our proposed MI sensor has good magnetic field tracking performance and is sensitive to the magnetic anomalies in the measured area.

\section{Conclusions}
\label{sect:conclusion}
In this paper, we describe a miniature MI sensor module with dimensions of 1 cm x 1 cm x 0.2 cm to measure the low vector magnetic field. The laboratory testing results demonstrate that the proposed MI sensor can detect a low magnetic field strength as low as 8.28 nT when the sampling rate is 10 Hz, the typical noise floor is about 0.235 nT/Hz$^{1/2}$ at 1 Hz, the sensitivity is about 13 nT, and the non-linearity is no more than 3\%. For the field testing results, the proposed MI sensor shows good magnetic field tracking performance, and has the ability to catch magnetic anomalies. Consequently, the proposed MI sensor shows a great scope for magnetic anomaly detection, and the miniature property makes it overcome the shortcomings of the related commercial devices (especially the fluxgate sensor) for measuring the low vector magnetic field, i.e., bulky size, heavy, high power, etc. Further, this study also lays the foundation for future works to improve the accuracy of the MI sensor for magnetic field detection.

\section*{Acknowledgment}
This work is partly supported by the Natural Science Foundation of Hubei Province of China under Grant No. 2020CFB610, the National Natural Science Foundation of China under Grant No. 41904164, the Foundation of Wuhan Science and Technology Bureau under Grant No. 2019010701011411, the Foundation of National Key Research and Development Program of China under Grant No. 2018YFC1503702, and the Fundamental Research Funds for the Central Universities, China University of Geosciences (Wuhan) under Grant No. CUG190628.

\section*{Data availability} 
The data that support the findings of this study are available from the corresponding author upon reasonable request.

\section*{References}
\nocite{*}
\bibliography{aipsamp}

\end{document}